# Modeling of Magnetic Properties of NiCl$_2$ Nanostripes, Nanotubes and Fullerenes

A.N. Enyashin, A.L. Ivanovskii[*]

*Institute of Solid State Chemistry, Ural Branch of the Russian Academy of Sciences, 620990 Ekaterinburg, Russia*

**A B S T R A C T**

We show that the magnetic properties of antiferromagnetic layered NiCl$_2$ can be altered under nano-sizing depending on dimensionality and morphology type of the corresponding nano-forms. By means of Monte-Carlo simulations within classical Heisenberg model, the spin ordering, magnetic part of heat capacity $C_v$ and Néel temperatures $T_N$ for multi-walled NiCl$_2$ nanotubes and a fullerene were calculated and analyzed in comparison with the bulk NiCl$_2$ and multilayered two-dimensional NiCl$_2$ crystals and nanostripes. We have found that the nano-structuring of NiCl$_2$ at the size reducing and at the formation of the closed structures can influence propitiously on the preservation of antiferromagnetic properties and the suppression of superparamagnetic limit.

## 1. Introduction

Among the magnetic materials suitable for a wide range of practical applications, quasi-two-dimensional antiferromagnetic systems have attracted interest. The well known representatives of these systems are dihalides of *d*-metals with CdCl$_2$ structural type [1], for example, nickel dichloride NiCl$_2$ [2]. In the ground state, NiCl$_2$ contains ferromagnetic layers of nickel ions, which are antiferromagnetically ordered among themselves and in a weak magnetic field undergo the transition from antiferromagnetic phase to ferromagnetic phase [3, 4].

---

[*] Corresponding author.
*E-mail address:* ivanovskii@ihim.uran.ru (A.L. Ivanovskii).

In the last five years appreciable results have been achieved in the preparation of a wide group of nanosized *d*-metal dihalides ($NiCl_2$, $FeCl_2$, $CdCl_2$, $CdI_2$, $NiBr_2$) with morphology of nanotubes, octahedral fullerenes and fullerene-like nanoparticles by laser, electron beam or thermal evaporation and condensation [5-11]. However, there are no reliable data on the electronic and magnetic properties of these nanostructures, because they appear as a mess of particles with different size and morphology, which are unstable against humid atmosphere. It prevents from a massive and expedient production of nanostructured halides and explains a poor characterization of hollow halide nanostructures to date.

Recent experiments on capillary imbibition of molten $PbI_2$ by $WS_2$ nanotubes have resulted in the preparation of core-shell $PbI_2@WS_2$ nanotubes [12], which shows a direction for guided production of protected and well crystallized halide nanotubes with more or less a uniform size, depending only on the size of the chalcogenide nanotubes [13]. The diameters of chalcogenide nanotubes are typically an order of magnitude larger than the diameters of carbon nanotubes [14], and the final product after capillary imbibition and cooling of molten $PbI_2$ within $WS_2$ nanotubes may differ from bulk nanocrystallites, since the strain energy of $PbI_2$ monolayer adsorbed on $WS_2$ surface will be not a limiting factor for a formation of rolled $PbI_2$ structures. The formation of multiwall nanotubes of layered $PbI_2$ coaxial to the $WS_2$ nanotubes can be easily observed [12, 15]. It opens new horizons in the guided fabrication of nanostructures of related halides, since the capillary filling of chalcogenide nanotubes can offer in a nearest future the production of protected and well crystallized halide nanostructures with more or less uniform size and permanent electronic and magnetic properties, depending only on the size distribution of the chalcogenide nanotubes [16].

In this work we present preliminary results about a possible influence of nanostructuring on the magnetic properties of $NiCl_2$. Using Monte-Carlo simulations within framework of classical Heisenberg model the spin ordering, magnetic part of heat capacity $C_v$ and Néel temperatures $T_N$ for multi-walled $NiCl_2$ nanotubes and a fullerene were calculated in comparison with the bulk NiCl$_2$ and corresponding multilayered two-dimensional NiCl$_2$ crystals and nanostripes.

**2. Structure models and computational method**

*2.1. Models of NiCl$_2$ nanostripes, nanotubes and fullerenes*

NiCl$_2$ adopt layered CdCl$_2$-type structure (space group *R-3m*, No 16, *Z* = 3), where each Ni ion is coordinated by six Cl ions, and each Cl ion is bonded to three Ni centers [3, 17]. For a description of antiferromagnetic state of the bulk NiCl$_2$ we have used rectangular supercell with the parameters *a* = 3.468 Å and *c* = 34.40 Å, which contained six layers. Such model was used also for the simulation of the bulk material and two-dimensional double-, triple- and six-layered NiCl$_2$ crystals and corresponding one-dimensional NiCl$_2$ nanostripes, which supercells contained in general 2880 Ni atoms.

Architecture of nanotubes, which are based on hexagonal NiCl$_2$ monolayer, can be characterized within framework of the same classification like for carbon and other related inorganic nanotubes [18, 19]. Among single-walled nanotubes three types of chirality can be distinguished: *zigzag* (*n*,0), *armchair* (*n*,*n*) and *chiral* (*n*,*m*), with *m* ≠ *n* and *m* ≠ 0. Here we consider multi-walled nanotubes as a system of coaxial single-walled nanotubes. In spite of many other layered compounds (like graphite, MoS$_2$, WS$_2$ etc.) the perfect matching between lattice parameters *a* and *c* of the bulk NiCl$_2$ gives an unique opportunity for the existence of nanotubes with the same interwall distance as van der Waals' gap in the bulk material, i.e., the difference between chiral indexes for two adjacent *armchair* nanotubes $\Delta n$ will have exact integer value calculated using relation $\Delta n = \frac{\pi}{3\sqrt{3}} \frac{c}{a} = 6.00$. In general, we have considered three multi-walled nanotubes with the same outer diameter 95.6 Å: double- (44,44)@(50,50), triple- (38,38)@(44,44)@(50,50) and six-walled

(20,20)@(26,26)@(32,32)@(38,38)@(44,44)@(50,50), which are composed by 10 unit cells and contained 1880, 2640 and 4200 Ni atoms, respectively.

The models for $NiCl_2$ fullerenes can be created in a similar way as for other inorganic layered compounds with $MX_2$ stoichiometry [20, 21]. This procedure includes the change in the evenness of M-X cycles. Namely, an introducing of six four-membered cycles (square-like defects) may result in the construction of octahedral particles as observed in experiments [5, 6]. Triangular facetes of these octahedral nanoparticles are composed by the fragments of $NiCl_2$ monolayer, while at the corners square-like defects are placed. In our work we have considered double-walled nanooctahedron $(NiCl_2)_{2304}@(NiCl_2)_{3136}$ containing 5440 Ni atoms.

*2.2. Computational details*

As a starting point of our method, classical Heisenbergs model has been applied. Hamiltonian for magnetic $NiCl_2$ in the absence of magnetic field may be written in the form [3]:

$$H = -\sum_{i>j} 2J_{ij} S_i \cdot S_j - \sum_{i>j'} 2J'_{ij'} S_i \cdot S_{j'} - D\sum_i S_i^{x2} - D'\sum_i \left(S_i^{y2} - S_i^{z2}\right), \quad (1)$$

where $J_{ij}$ and $J'_{ij}$ are exchange integrals for inter- and intralayer spin interactions ($J_{ij}$ is equal 21.70 K and -4.85 K for the first and second coordination spheres, respectively, $J'_{ij}$ = - 0.77 K), $D$ and $D'$ are constants of plane anisotropy ($D = D' = 0.40$ K), $S_i \equiv (S_i^x, S_i^y, S_i^z)$ is classical spin (with modulus equal 1). Aforementioned fact of perfect matching between $a$ and $c$ lattice parameters and equality of interlayer distances as well as intralayer interatomic distances both in the bulk and nanotubular $NiCl_2$ allow to use the same values of $J_{ij}$ and $J'_{ij}$ without a considerable correction for the distance between the spins.

Thermodynamical properties of crystalline and nanostructured allotropes of $NiCl_2$ were calculated using Monte-Carlo method (MC) applying Metropolis algorithm [22]. For every spin $5 \cdot 10^4$ MC steps were performed, at that first $3 \cdot 10^4$ MC steps were used for thermalization, while the rest steps were intended for the calculation of magnetic part in molar heat capacity $C_v$ using statistical fluctuations of total energy $E$ at temperature $T$ as:

$$C_v = \frac{1}{kT^2}\left(\langle E^2 \rangle - \langle E \rangle^2\right). \qquad (2)$$

## 3. Results and discussion

First, the simulation was performed for the bulk $NiCl_2$ in order to estimate an error in the calculation of the magnetic ordering temperature $T_N$ (Néel temperature). The calculated dependence of magnetic part of molar heat capacity $C_v$ on temperature $T$ demonstrates clear peak at $T = 63$ K which corresponds to a phase transition (Fig. 2) and may be attributed to $T_N$. An analysis of the spin ordering depending on the temperature proves this supposition (Fig. 3). At T > 63 K the spin ordering is lost, and the magnetic structure of a layer within the bulk is represented by the "islands" of Ni ions with different spin orientation, while at T < 63 K the magnetic structure is represented as a classical antiferromagnetic system composed by layers with alternating spin orientation along $c$-axis. Thus, the use of the above mentioned values of exchange integrals for inter- and intralayer spin interactions [3] leads to an overestimation of $T_N$ on ~ 10.7 K in comparison with experimental data $T_N = 52.3$ K [3]. Therefore, their use in further simulations of magnetic properties of $NiCl_2$ systems can provide a semi-quantitative estimation.

Let us note that the origin of second low-intensity peak of $C_v$ at $T = 25$ K is not clear and its existence has not been reported in the past [2]. Based on the results of our MC simulations, we may suppose that this peak is related to partial low-temperature change of the spin orientations: at temperatures below 25 K a kind of spin waves is obtained across $c$-axis of the crystal, which disappears at higher temperatures. Though, this phenomenon may be an artifact of the model applied, when classical Heisenbergs Hamiltonian (1) fails in description of thermodynamical properties at low temperatures, and quantum effects should be accounted.

Simulations for two-dimensional crystals of $NiCl_2$ (the stacks composed by two, three and four of infinite $NiCl_2$ layers) reveal that antiferromagnetic ordering for such systems is preserved. The dependence of $C_v$ on $T$ also have a pronounced peak near ~60 K (Fig. 2). However, the values $T_N$ comparing with that of the bulk crystal are lower. For example, $T_N$ for $NiCl_2$ double layer is increased on 6 K comparing to the bulk (Table 1). This phenomenon can be explained by the absence of a part of interaction of the surface spins - antiferromagnetic coupling with the spins of adjacent layer, which facilitates their easier disordering at the temperature growth. With increasing of the number of layers, the $T_N$ values of two-dimensional $NiCl_2$ approach the $T_N$ of the bulk. An easier ability for rotation of the surface spins and for their easier disordering can be supported by the calculations of the energies of spin ordering $ΔE$ (Table 1), which also decrease when the number of layers increases.

Similar behavior is observed for one-dimensional planar nanostripes of $NiCl_2$ (Fig. 2). All nanostripes keep antiferromagnetic ordering, but their Néel temperatures become lower comparing even to the infinite two-dimensional crystals with corresponding number of the layers (Table 1). This decreasing $T_N$ can be explained by the presence of the spins at the edges of nanostripes, which have smaller coordination number, therefore, a weaker ferromagnetic coupling with the spins within the same layer.

Thus, simple nanosizing does not change considerably antiferromagnetic properties of $NiCl_2$, but causes their slight decaying – owing to an easier ability for rotation of the "surface" spins for two-dimensional $NiCl_2$ or for both "surface" and edges' spins of $NiCl_2$ nanostripes.

One may expect that this decaying can be partially prevented in the case of closed multiwalled nanostructures (such as coaxial nanotubes and fullerenes), since they would not posses the edges with the atoms of low coordination number like in nanostripes. However, these low-dimensional nanostructures have different number of spins in adjacent walls, i.e. the number of spins in one layer is not compensated by the spins of a neighboring layer. The influence of this genuine feature of multiwalled inorganic nanostructures is not evident *a priori*.

The results for MC simulations of magnetic ordering within multiwalled nanotubes are depicted in Fig 2 and 3 and listed in Table 1. All nanotubes have the dependences of $C_v$ on $T$, which are typical for above considered antiferromagnetics $NiCl_2$ structures. Though, already for double-walled $NiCl_2$ nanotubes no remarkable decrease in $T_N$ comparing to the bulk is observed – as against of nanostripes or two-dimensional crystals. For example, maximal deviation (from the bulk) $\Delta T_N = 2$ K was found for double-walled (20,20)@(26,26) nanotube, which is less than for a double-walled nanostripe with $\Delta T_N = 8$ K. Moreover, the nanotubes with number of wall more than two demonstrate even an increase $T_N$ on a few Ks (Table 1). High values of $T_N$ for multiwalled $NiCl_2$ nanotubes comparing with that of the bulk also correlate with relative energies of spin ordering $\Delta E$ (Table 1). In these cases they have negative values evidencing about stronger coupling of the spins. Antiferromagnetic ordering of spins within the walls of multiwalled $NiCl_2$ nanotubes is preserved until temperatures $T_N$ and no special feature was found in comparison with that of the bulk crystal (Fig. 3).

Finally, according to our simulations (Fig. 2), zero-dimensional double-walled fullerene $(NiCl_2)_{2304}@(NiCl_2)_{3136}$ is also an antiferromagnetic particle (Fig. 2.1). Like for $NiCl_2$ nanotubes a shift of $T_N$ in the area of higher temperatures and negative energy of spin ordering are found (Table 1).

## 4. Conclusions

Many properties of the solids undergo a considerable change or novel ones appear, when the size of particles is reduced. A special interest is given to the research of ferromagnetic properties of iron-group metal nanoparticles as materials for high-density recording [23]. Under certain limit the magnetic moment of these nanoparticles becomes unstable, and these nanoparticles acquire superparamagnetism. This process unwanted for applied purposes can be beat up using magnetic exchange coupling induced at the interface between ferromagnetic and antiferromagnetic systems [24]. However, a change of antiferromagnetic properties due to a nanosizing could be also possible, which can prevent the usage of exchange bias.

Among antiferromagnetic materials suitable for a range of practical applications *d*-metals dihalides attract a valuable interest. In this work atomic models of a set of low-dimensional $NiCl_2$ nanostructures – multilayered nanostripes, nanotubes and a fullerene - were constructed. Together with the bulk and two-dimensional crystals of $NiCl_2$ they were used to study an influence of dimensionality and morphology type on magnetic properties of $NiCl_2$ – a classical layered antiferromagnetic system.

Our results show that, all nanostructures considered remain antiferromagnetic. For planar two-dimensional $NiCl_2$ crystals and one-dimensional nanostripes antiferromagnetic state becomes less favorable, and the values of Néel temperature $T_N$ are shifted down comparing to that of the bulk. On the contrary, for curved and closed nanostructures - multi-walled nanotubes and fullerenes - a gain in energy of spin ordering testifies about slightly stronger spin interaction. Magnetic disordering in these hollow low-dimensional nanostructures appears at higher $T_N$, than for the bulk. Hence, a nanostructuring at the size reducing can have a propitious influence on the preservation of antiferromagnetic properties and the suppression of superparamagnetic limit.

**Acknowledgements**
The authors are grateful to the Russian Foundation for Basic Research (grant 10-03-96004-Ural).

**Table 1.**

Néel temperatures $T_N$ and relative energy of spin ordering $\Delta E$ within the bulk and nanostructured $NiCl_2$ depending on the dimensionality $D$ and morphology*

| System | $D$ | $T_N$, K | $\Delta E$, K/spin |
|---|---|---|---|
| bulk | 3 | 63 | 0 |
| infinite double layer | 2 | 57 | 3.11 |
| infinite triple layer | 2 | 60 | 2.34 |
| infinite six layers | 2 | 61 | 1.57 |
| doublelayered nanostripe | 1 | 55 | 4.37 |
| triplelayered nanostripe | 1 | 58 | 3.67 |
| six-layered nanostripe | 1 | 61 | 2.92 |
| 20@26 nanotube | 1 | 61 | -0.26 |
| 44@50 nanotube | 1 | 62 | -0.37 |
| 38@44@50 nanotube | 1 | 65 | -2.15 |
| 20@26@32@38 nanotube | 1 | 66 | -3.07 |
| 20@26@32@38@44@50 nanotube | 1 | 69 | -4.06 |
| $(NiCl_2)_{2304}$@$(NiCl_2)_{3136}$ fullerene | 0 | 68 | -7.07 |

* all nanotubes considered are of *armchair* chirality, only one index is indicated

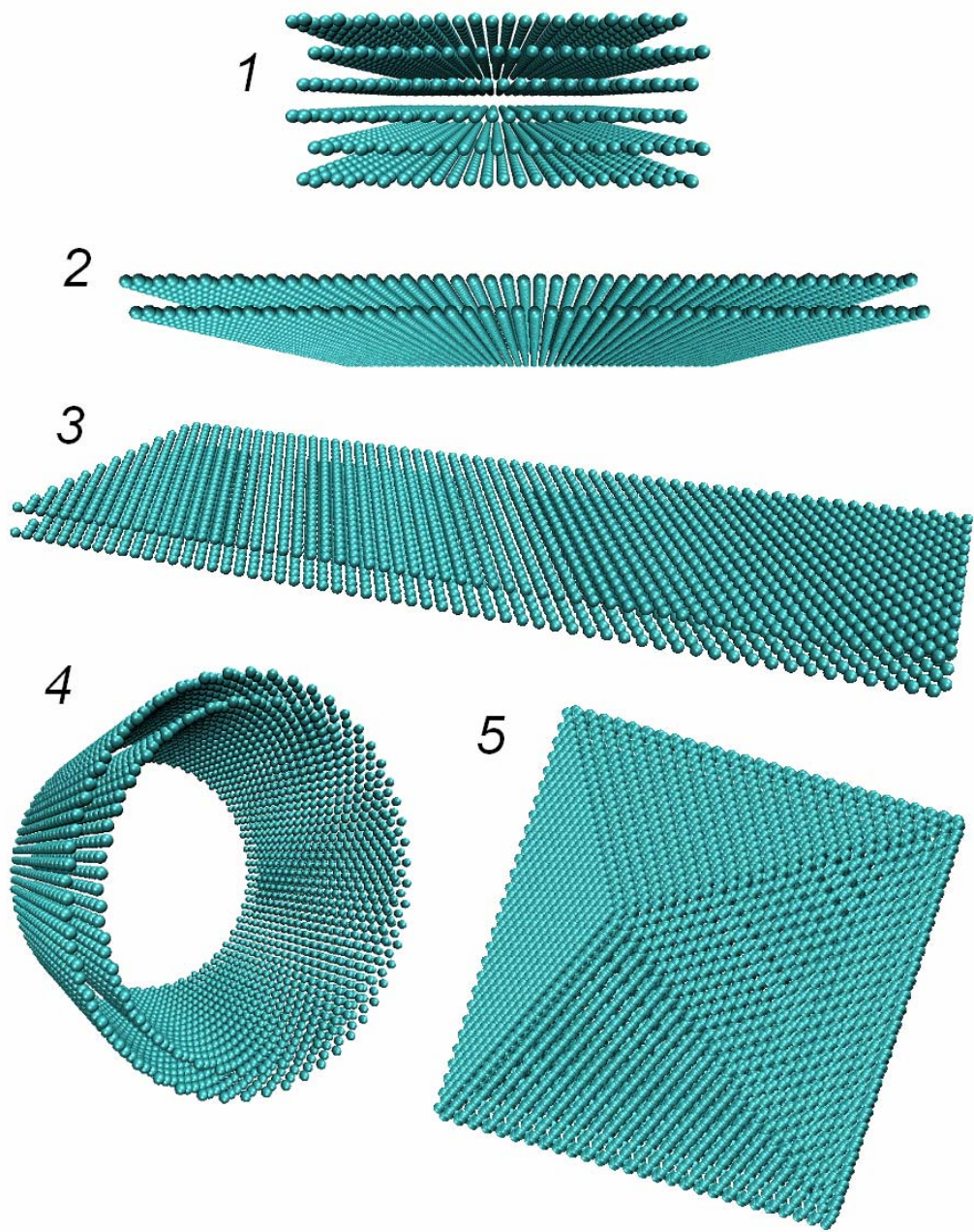

**Figure 1.** Structures of the bulk NiCl$_2$ (1), infinite NiCl$_2$ bilayer (2), one-dimensional (12,12) NiCl$_2$ nanostripe (3), double-walled (44,44)@(50,50) NiCl$_2$ nanotube (4) and fullerene (NiCl$_2$)$_{2304}$@(NiCl$_2$)$_{3136}$ (5). Nickel atoms are only shown.

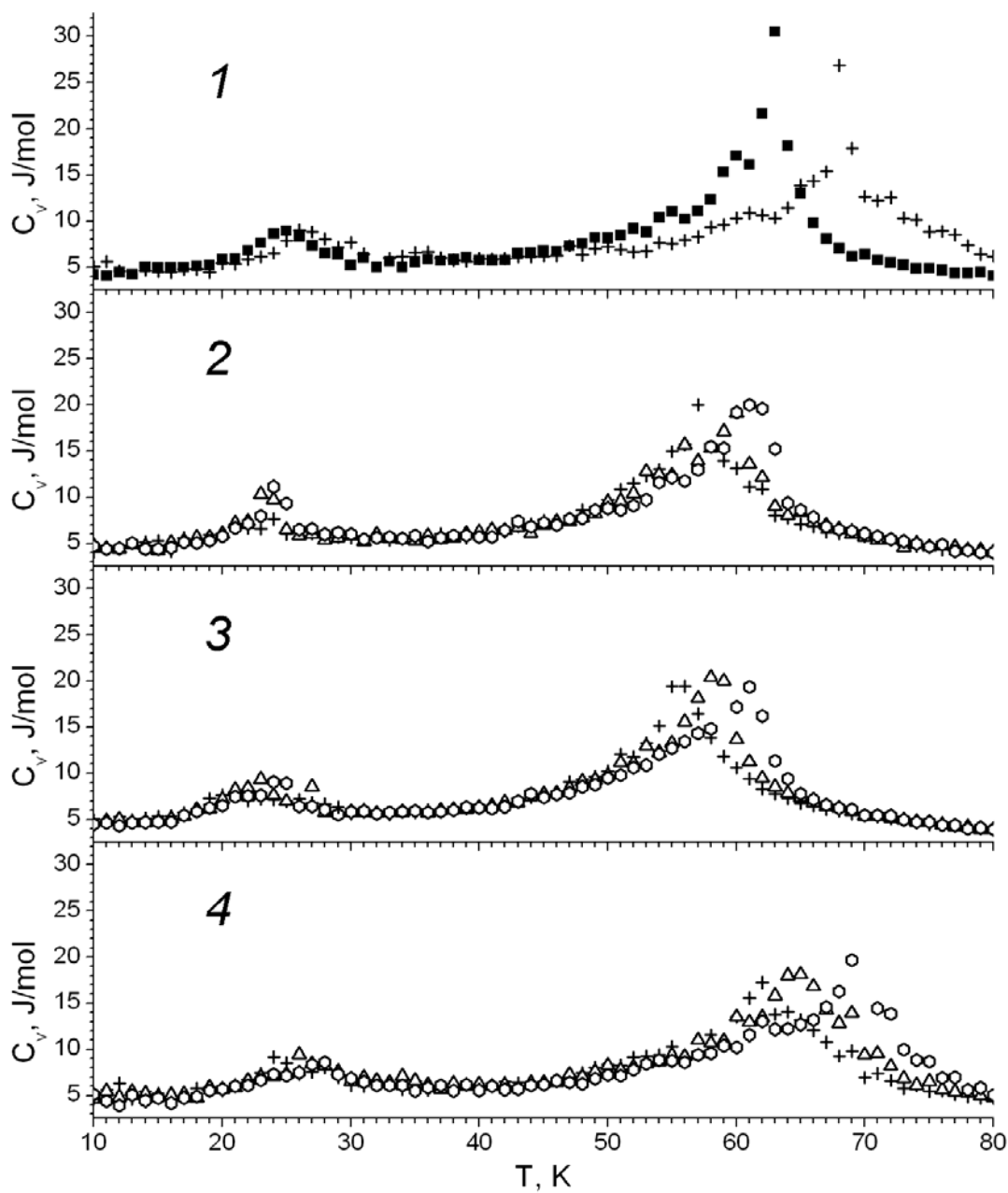

**Figure 2.** Calculated magnetic part in molar heat capacity for various nickel chloride allotropes depending on the temperature: *1* – the bulk NiCl$_2$ (■) and fullerene (NiCl$_2$)$_{2304}$@(NiCl$_2$)$_{3136}$ (+), *2* – infinite two-dimensional NiCl$_2$ crystals and *3* – NiCl$_2$ nanostripes containing two, three and six NiCl$_2$ layers (+, Δ, ○, respectively), and *4* – NiCl$_2$ nanotubes with chiralities (44,44)@(50,50) (+), (38,38)@(44,44)@(50,50) (Δ) and (20,20)@(26,26)@(32,32)@(38,38)@(44,44)@(50,50) (○)

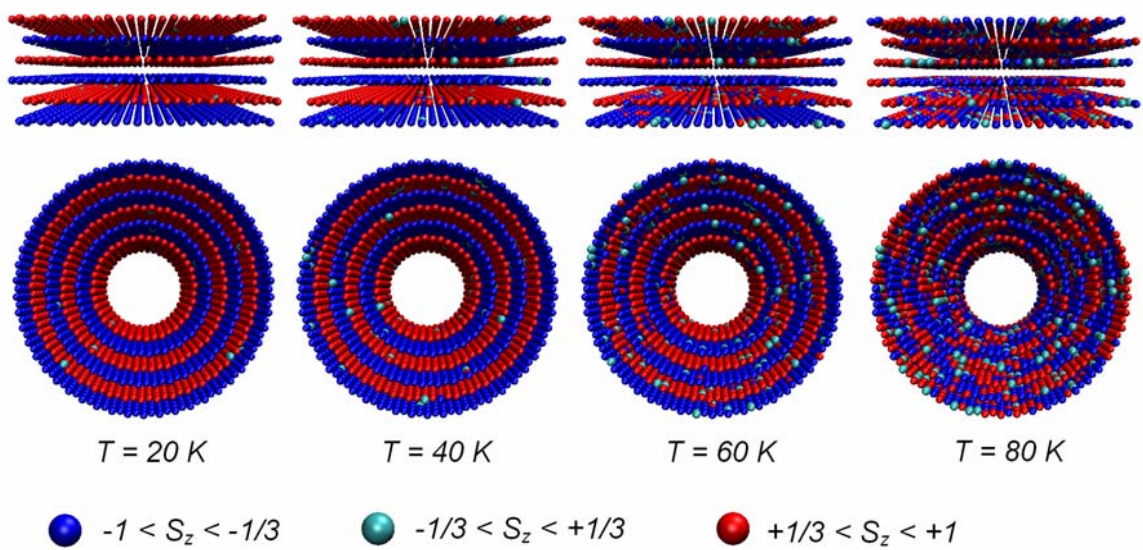

**Figure 3.** Snapshots of spin ordering in the bulk NiCl$_2$ and six-walled NiCl$_2$ nanotube (20,20)@(26,26)@(32,32)@(38,38)@(44,44)@(50,50) at different temperatures.